\documentstyle[prb,eqsecnum,aps]{revtex}
\begin{document}
\title{Finite Size  Analysis of the Structure Factors in the 
       Antiferromagnetic XXZ Model}

\author{M. Karbach, K.-H. M\"utter\cite{muetter} and M. Schmidt}
\address{Physics Department, University of Wuppertal\\ 
42097 Wuppertal, Germany}

\date{23.02.94}

\maketitle
\begin{abstract}
We perform a finite size analysis of the longitudinal and transverse structure
factors $S_j(p,\gamma,N),j=1,3$ in the groundstate of the spin-$\frac{1}{2}$ 
XXZ model. Comparison with the exact results of Tonegawa 
for the XX model yields excellent agreement. Comparison with the conjecture 
of M\"uller, Thomas, Puga and Beck  reveals discrepancies 
in the momentum dependence of the longitudinal structure factors.

\end{abstract}
\draft
%
%
%
%
\section{Introduction}
In a previous publication, \cite{karbach} two of us (M.K. and K.-H.M.) have 
studied the structure factor $S(p,N)$ in the groundstate of the 
antiferromagnetic Heisenberg (AFH)-model on finite rings with $N=4-30$ sites. 
We found that finite size effects die out with $N^{-2}$ and with a coefficient 
$c(p)$ increasing with momentum $p$. The ``gross''-structure of $S(p,N)$
is well described by $-\ln(1-p/\pi)$ for all values $p<\pi$.
Therefore indications for a critical behavior in the limit 
$p\rightarrow \pi$ can be seen already for all values of $p$. For this
reason the structure factor appears to be more suited then the spin-spin 
correlators to extract the critical behavior from finite systems.

In the present paper we are going to extend this analysis to the
XXZ-model with Hamiltonian:
\begin{equation}
	H=2\sum_{x=1}^{N}\bigl[ 
		S_1(x)S_1(x+1)+S_2(x)S_2(x+1)+
		\cos \gamma \; S_3(x)S_3(x+1)     \bigr],
\end{equation}
which coincides with the AFH-Hamiltonian in the isotropic limit $\gamma=0$.
The groundstate properties of the spin-spin correlators at large
distances $x\rightarrow \infty$ and in the critical regime 
$0 \leq \gamma \leq \pi/2$ have been found by Luther and Peschel\cite{luther}
\begin{equation}\label{ce}
	\langle 0|S_j(0)S_j(x)|0\rangle \stackrel{x\to\infty}{\longrightarrow} 
		\frac{(-1)^x}{x^{\eta_j}} ,\quad j=1,3,
\end{equation}
with crititcal exponents:
\begin{equation}\label{sc}
	\eta_1=\eta_3^{-1}=1-\frac{\gamma}{\pi}, 
		\quad {\rm for} \quad 0 \leq \gamma \leq \frac{\pi}{2}. 
\end{equation}
The critical exponent $\eta_3$ were derived also from the quantum inverse 
scattering method.\cite{korepin} Both exponents $\eta_1$ and $\eta_3$ are also 
predicted by conformal invariance.\cite{ci}

The structure factors are defined as the Fourier transforms of the 
longitudinal and transverse spin-spin correlators:
\begin{eqnarray}
	S_j\left(p=2\pi k/N,\gamma,N\right)
	&=&1+(-1)^k 4\cdot\langle 0|S_j(0) S_j(N/2)|0\rangle \\
	&&+8\cdot\sum_{x=1}^{N/2-1}\langle 0|S_j(0)S_j(x)|0\rangle
	\cdot\cos (px) ,\quad j=1,3. \nonumber
\end{eqnarray}
The large distance properties of the spin-spin correlators (\ref{ce}) 
induce a specific behavior of the corresponding structure factors
for momenta $p\rightarrow \pi$:
\begin{equation}\label{sptopig}
	S_j(p,\gamma)=a_j+b_j\cdot\left(1-\frac{p}{\pi}\right)^{\eta_j-1}
	\quad {\rm for} \quad j=1,3 \quad \text{ and }
	\quad 0 < \gamma \leq \frac{\pi}{2},
\end{equation}
where the critical exponents $\eta_j, j=1,3$ are given by (\ref{sc}).
Note that the longitudinal structure factor stays finite for
$p=\pi$ and $0 < \gamma \leq\pi/2$, whereas
the transverse structure factor diverges. The questions we adress here
are the following: 

\begin{itemize}
\item[(1)] Is it possible to see the specific behavior (\ref{sptopig})
	already on finite systems?
\item[(2)] Is rotation invariance in spin space restored in the isotropic
	limit $\gamma\to 0, \eta_j \rightarrow 1$?
\item[(3)] What is the leading singularity of $S_j(p,\gamma=0,N=\infty)$
		for $p \to \pi$?
\item[(4)] What is the leading large $N$ behavior of the structure factors
	$S_j(p=\pi,\gamma,N)$ in the isotropic limit $\gamma\to 0$?
\end{itemize}
The answer to the second question is of special interest in view of
the results of Singh, Fisher and Shankar.\cite{fisher} Following 
Ref.\onlinecite{fisher}, the transverse and longitudinal structure factors 
$S_j(p=\pi,\gamma',N=\infty),j=1,3$ behave differently if we approach the 
isotropic limit from the noncritical regime $\cos \gamma=\cosh \gamma' >1$:
\begin{equation}\label{sjppiginf}
	S_j(p=\pi,\gamma',N=\infty)\rightarrow \gamma'^{-\lambda_j},
\end{equation}
where
\begin{equation}\label{l1l2}
	\lambda_1=1.5, \quad \lambda_3=2.
\end{equation} 
It is well known \cite{johnson} that in the noncritical regime 
$\gamma' > 0$ the large distance behavior of the correlators is governed by a
finite correlation length:
\begin{equation}
	\xi(\gamma') \stackrel{\gamma'\to 0}{\longrightarrow} 
	 \frac{1}{8} \exp\left(\frac{\pi^2}{2\gamma'}\right),
\end{equation}
which diverges exponentially for $\gamma'\to 0$. Therefore, both longitudinal
and transverse structure factors stay finite at $p=\pi, \gamma'\neq0$. In order
to see the effect of the divergencies (\ref{sjppiginf}) and (\ref{l1l2}) on 
finite rings, one could think of a finite size scaling ansatz of the 
structure factors: 
\begin{equation}\label{sjpgngj}
	S_j(p=\pi,\gamma',N) = S_j(p=\pi,\gamma',N=\infty) \cdot g_j(z),
\end{equation}
in the combined limit
\begin{equation}\label{ntoinf}
	N\to\infty,\quad\gamma'\to 0,\quad z = \frac{N}{\xi(\gamma')}
		\quad {\rm fixed}.
\end{equation}
Note, however, that the correlation length increases exponentially in the limit
$\gamma' \to 0$. Therefore, extremely large systems are needed, to extract the
structure factors $S_j(p=\pi,\gamma',N=\infty)$ close to the isotropic limit.
Having this in mind, we did not make an attempt, to see the effect of the
divergencies (\ref{sjppiginf}) on our rings with $N\leq 28$ sites. 

The outline of the paper is as follows. In Sec. II we demonstrate, that the 
longitudinal and transverse structure factors obey approximate scaling laws 
for all momentum values $0 < p < \pi$ and $\gamma$-values 
$0\leq \gamma \leq \pi/2$. In Sec. III, we perform a finite size
analysis of the structure factors for noncritical momenta 
$p \leq 6\pi/7$. In Sec. IV we study how the transverse and
longitudinal structure factors at $p=\pi$, converge to each other in the limit
$\gamma \to 0$. Finally, we propose in Sec. V a finite size scaling ansatz with
an appropriate scaling variable, which accounts for all finite size effects at
noncritical momenta $p < \pi$ and at the critical momentum $p=\pi$.
\section{Approximate scaling properties of the longitudinal and transverse 
structure factors}
In Fig. 1 we have plotted the longitudinal
structure factor $S_3(p,\gamma,N)$ on finite rings with $N=4,6,...,28$
sites and $\gamma/\pi=0,0.1,0.2,0.3,0.4,0.5$
as function of the ``scaling variable'':
\begin{equation}\label{l3pg}
	L_3(p,\gamma)=\frac{\eta_3}{\eta_3-1}
		\left[1-\left(1-\frac{p}{\pi}\right)^{\eta_3-1}\right].
\end{equation}
The normalization in the scaling variable (\ref{l3pg}) has been chosen
such that the limit to the isotropic case $\eta_3 \rightarrow 1$
can be performed:
\begin{equation}\label{l3pg0}
       \left.L_3(p,\gamma)\right|_{\gamma=0}=-\ln\left(1-\frac{p}{\pi}\right).
\end{equation}
Indeed (\ref{l3pg0}) has been found in Ref.\onlinecite{karbach}  as the 
adequate  variable in the isotropic case. Looking at Fig. 1, we observe that 
all data points follow approximately a single straight line with slope one as 
long as $p<\pi$. For these momentum values finite size effects appear to be 
small in contrast to the ``critical'' point $p=\pi$, where finite size effects 
are large. Deviations from the straight line in Fig. 1 and finite size effects
become visible in Fig. 2 where we have plotted the difference:
\begin{equation}\label{d3pgn}
	\Delta_3(p,\gamma,N)=S_3(p,\gamma,N)-L_3(p,\gamma),
\end{equation}
versus the scaling variable (\ref{l3pg}). Note the surprising behavior of the
longitudinal structure factor of the XX-model:
\begin{equation}
	S_3(p,\gamma=\pi/2,N)=L_3(p,\gamma=\pi/2)=\frac{2}{\pi}p.
\end{equation}
It is free of any finite size effect! This is proven in Appendix A 
using the rigorous
results of Lieb, Schultz and  Mattis. \cite{lsm} For decreasing values of
$\gamma$ we observe in the differences (\ref{d3pgn}) a ``fine'' structure with a 
minimum for larger $p$-values. Moreover, there seems to be a maximum
for small $p$ and $\gamma $ values.

Let us now turn to the transverse structure factor $S_1(p, \gamma,N)$
which we present in Fig. 3 as function of the analogous ``transverse
scaling variable'':
\begin{equation}\label{l1pg}
	L_1(p,\gamma)=\frac{\eta_1}{\eta_1-1}
		\left[1-\left(1-\frac{p}{\pi}\right)^{\eta_1-1}\right].
\end{equation}
Both variables (\ref{l3pg}) and (\ref{l1pg}) coincide in the isotropic 
limit (\ref{l3pg0}).
Again we observe an approximate scaling, which however is less impressiv
than the scaling of the longitudinal structure factor. In order to
resolve detailed features and finite size effects we plot again in
Fig. 4 the difference:
\begin{equation}\label{d1pgn}
	\Delta_1(p,\gamma,N)=S_1(p,\gamma,N)-L_1(p,\gamma),
\end{equation}
as function of $L_1$. A comparison of Figs. 2 and 4 reveals two important
differences in the finite size effects of the longitudinal and
transverse structure factors $S_3$ and $S_1$:  
\begin{itemize}
\item[(1)] They are zero in $S_3(p,\gamma =\pi/2,N)$, but nonzero in 
	$S_1(p,\gamma =\pi/2,N)$.
\item[(2)] They are small in $S_3$ but large in $S_1$ for small momentum 
	values.
\end{itemize}
\section{Finite size analysis of the  structure factors for noncritical
momenta}
We have analyzed the finite size dependence of the 
structure factors for fixed momenta:
\begin{eqnarray}
	p = \left\{
	\begin{array}{ll}
	\frac{\pi}{4},\frac{3\pi}{4} & \text{ for } N=8,16,24 \\ \nonumber
	\frac{\pi}{3},\frac{2\pi}{3} & \text{ for } N=6,12,18,24  \\ \nonumber
	\frac{\pi}{2} & \text{ for } N=4,8,12,16,20,24,28,
	\end{array}\right. 
\end{eqnarray}
which can be realized on the systems with size $N$ given above.
The $N$-dependence of the differences (\ref{d3pgn}) and (\ref{d1pgn}) can be 
parametrized by:
\begin{equation}\label{djpgn}
	\Delta_j(p,\gamma,N)=\Delta_j(p,\gamma)+\frac{c_j(p,\gamma)}{N^2},
		\quad j=1,3.
\end{equation}
As an example we show in Figs. 5(a) and 5(b) the finite size effects of the
structure factors at:
\begin{eqnarray}\nonumber
	p=\frac{\pi}{2} \text{ and }
	 \frac{\gamma}{\pi}=0,0.1,0.2,0.3,0.4,0.5.
\end{eqnarray}

The slopes $c_j(p,\gamma)$ turn out to be small for small 
momentum values $p<\pi/2$
and increase rapidly for $p \rightarrow \pi$. Indeed there seems to be a
singularity of the type $(1-p/\pi)^{\eta_j-3}$. 
Therefore we have plotted in Fig. 6 the slopes versus the variable:
\begin{equation}\label{varfig6}
	\frac{\eta_j(\eta_j-2)}{2} 
	\left[1-\left(1-\frac{p}{\pi}\right)^{\eta_j-3}\right].
\end{equation}
We observe an almost linear behavior in this variable. Note that Fig. 6
allows us to compare the finite size effects (\ref{djpgn}) of the longitudinal
and transverse structure factors: They coincide in the isotropic limit 
but differ more and more with increasing $\gamma$. A motivation for
the special choice of the variable (\ref{varfig6}) is given in the  
Sec. V. In Fig. 6 we have included as well the momenta:
\begin{eqnarray}\label{ppimq2dn}
	p = \left\{
	\begin{array}{ll}
	\frac{\pi}{5},\frac{2\pi}{5},\frac{3\pi}{5},\frac{4\pi}{5} 
		& \text{ for } N=10,20 \nonumber \\ \nonumber
	\frac{\pi}{6},\frac{5\pi}{6} & \text{ for }  N=12,24 \\ \nonumber
	\frac{\pi}{7},\frac{2\pi}{7},\frac{3\pi}{7},\frac{4\pi}{7},
		\frac{5\pi}{7},\frac{6\pi}{7} & \text{ for }  N=14,28,
	\end{array}\right.
\end{eqnarray}
which occur on  two systems. Here we have assumed that the finite size 
dependence is described correctly by (\ref{djpgn}). Our estimates for the 
structure factors in the thermodynamical limit can be seen from the solid 
dots and squares in Figs. 7(a)-(f).

So far our estimate of the thermodynamical limit is restricted to
momentum values $p\leq 6\pi/7$ due to the finiteness of our
systems $N\leq 28$. Additional information on the structure factors
for $p$-values:
\begin{equation}\label{ppim2dn}
	p=\pi\left(1-\frac{2}{N}\right), \quad N=14,16,...,28,
\end{equation}
close to the ``critical'' momentum $p=\pi$ can be obtained, if finite
size scaling holds for the structure factors
\begin{equation}\label{sa}
	S_j(p,\gamma,N)=S_j(p,\gamma,N=\infty)\cdot g_j\left(z,\gamma\right),
\end{equation} 
in the combined limit
\begin{equation}
	N\rightarrow \infty,\quad p\rightarrow \pi, 
	\quad z=\left(1-\frac{p}{\pi}\right) N \quad {\rm fixed}.
\end{equation}
Note, that the scaling variable $z$ is just $2$ for the
momenta listed in (\ref{ppim2dn}). The scaling function at this value can be
taken from:
\begin{equation}
	g_j(2,\gamma)=\frac{S_j(p=6\pi/7,\gamma,N=14)}
		           {S_j(p=6\pi/7,\gamma,N=\infty)},
\end{equation}
and is shown as function of $\gamma$ in Fig. 8.
In this way we get from the finite size scaling ansatz (\ref{sa}) an estimate
of the thermodynamical limit of the structure factors:
\begin{equation}
	S_j \biglb(p=\pi(1-2/N),\gamma,\infty \bigrb)=
	\frac{S_j\biglb(p=\pi(1-2/N),\gamma,N \bigrb)}{g_j(2,\gamma)}
\end{equation}
for the momenta listed in (\ref{ppimq2dn}). The result for the 
differences (\ref{d3pgn}) and (\ref{d1pgn}) is marked in Figs. 7(a)-(f) 
by the open dots and squares.

The validity of our finite size approach to the structure factors in the
thermodynamical limit can be tested in case of the XX model ($\gamma = \pi/2$),
where the transverse structure factor is known exactly from the results of
Tonegawa.\cite{tonegawa} 

The dotted curve in Fig. 7(a) represents the exact result; the solid and open
squares the results of our finite size analysis.

In Ref. \onlinecite{mueller} M\"uller {\it et al.} have made a conjecture on
the dynamical correlation functions in the XXZ model. From this conjecture one
gets a prediction on the static structure factors 
$S_1(p,\gamma), S_3(p,\gamma)$, which we have plotted in Figs. 7(a)-(f) by the
dashed and solid curves, respectively. We observe:
\begin{itemize}
	\item[(1)] Excellent agreement of our finite size analysis with the
		exact result of Ref. \onlinecite{tonegawa} for $\gamma=\pi/2$.
		 The conjecture of M\"uller \cite{mueller} {\it et al.}  for 
		the transverse structure factor is systematically above the 
		exact result.
	\item[(2)] The conjecture of M\"uller\cite{mueller} {\it et al.} for
		the transverse structure factor at 
		$\gamma/\pi = 0.4, 0.3, 0.2, 0.1, 0$ deviates from our finite 
		size analysis more and more, if the momentum increases and if 
		$\gamma$ decreases.
	\item[(3)] The conjecture of M\"uller \cite{mueller} {\it et al.} for
		the longitudinal structure factor at 
		$\gamma/\pi = 0.4, 0.3, 0.2, 0.1, 0$ is in striking 
		disagreement from our finite size analysis. In particular, the 
		minimum we find in the difference 
		$\Delta_3(p,\gamma)=S_3(p,\gamma)-L_3(p,\gamma)$ at momentum 
		$p=p(\gamma)$ listed in TABLE I:

			\begin{table}
			\caption{The position $p=p(\gamma)$ of the minimum of 
				$\Delta_3(p,\gamma)$ 
			 as a function of $p(\gamma)$.}
			\begin{tabular}{cccccc}
			$\gamma/\pi$&$0.4$&$0.3$&$0.2$&$0.1$&0\\ \hline
			 $p(\gamma)/\pi$ &0.51 &0.59 &0.66 &0.73 &0.82 \\
			\end{tabular}
			\end{table}	

		is not predicted by the conjecture of Ref. 
		\onlinecite{mueller}. Beyond the minimum our finite size 
		analysis yields an increase of the difference (\ref{d3pgn}) 
		which might be linear or even stronger in the scaling variable
		$L_3(p,\gamma)$ [cf.(\ref{l3pg})]. A linear extrapolation for 
		the structure factor in the isotropic case yields:
		\begin{eqnarray}\label{s3pgna}
			\Delta_3(p,\gamma=0) = S_3(p,\gamma=0) + 
			   \ln \left(1-\frac{p}{\pi}\right) =
			- A \cdot \ln \left(1 -\frac{p}{\pi}\right)
			\qquad \text{ for } p \longrightarrow \pi,
		\end{eqnarray}
		where 
		\begin{eqnarray}
			A \gtrsim 0.12 .
		\end{eqnarray}
		We can also extrapolate to the critical momentum, assuming,
		that the prediction of renormalization group equations 
		\cite{fisher,affleck} is correct:
		\begin{eqnarray}\label{d3pgna}
			\Delta_3(p,\gamma=0) = S_3(p,\gamma=0) + 
			   \ln \left(1-\frac{p}{\pi}\right) = 
			   B \cdot
			   \left[
			    -\ln \left(1-\frac{p}{\pi}\right)
				\right]^{3/2}
		         \qquad \text{ for } p \longrightarrow \pi.
		\end{eqnarray}
		In this case we obtain for coefficient $B$:
		\begin{eqnarray}
			   B  \simeq 0.05.
		\end{eqnarray}
		Of course the vicinity of the minimum at $p=p(\gamma)=0.82\pi$,
		prevents us to decide, whether (\ref{s3pgna}) or (\ref{d3pgna})
		is the correct form at the critical momentum $p=\pi$.
\end{itemize}
\section{Finite size analysis of the structure factors at the critical 
         momentum }
The finite size ansatz (\ref{djpgn}) breaks down for the ``critical'' 
momentum $p=\pi$.
The exponent of $N^{-1}$ changes from a fixed to an $\eta_j$ dependent
value:
\begin{equation}\label{sjppign}
	S_j(p=\pi,\gamma,N)=r_j(\gamma) \cdot \frac{\eta_j}{\eta_j-1}
 	\left[1-\left(\frac{N}{N_j(\gamma)}\right)^{1-\eta_j}+...\right].
\end{equation}
We have plotted in Fig. 9 the longitudinal and transverse structure
factors at the critical momentum $p=\pi$ versus:
\begin{equation}\label{varppi}
	\frac{\eta_j}{\eta_j-1}
	\left[1-\sin \left(\frac{\pi}{2}\eta_j\right)
		\cdot N^{1-\eta_j}\right], 		\quad j=1,3.
\end{equation}
We observe an almost linear behavior in this variable. A tiny 
nonlinearity appears in the longitudinal structure factor if we approach the 
isotropic limit $\gamma \rightarrow 0$, where (\ref{varppi}) reduces to 
$\ln N$. The variable (\ref{varppi}) enables us to compare the longitudinal 
and transverse structure factors at momentum $p=\pi$. In particular, we can see
how the two structure factors converge to each other in the
isotropic limit. In this limit the coefficient in front of the
right hand side of (\ref{sjppign}) develops a pole, such that we find for the
structure factors: 
\begin{equation}\label{sjppig0n}
	S_j(p=\pi,\gamma=0,N) \stackrel{N\to\infty}{\longrightarrow}
		 r_j(0)\ln \left( \frac{N}{N_j(0)} \right).
\end{equation} 
For a precise determination of $r_1(0),r_3(0),N_1(0),N_3(0)$ we have repeated 
the computation of the structure factors for small $\gamma$ values
($\gamma/\pi=0.01,0.02,0.03,0.04,0.05$). The fit to the  ansatz
(\ref{sjppign}) yields for the parameters $r_j(\gamma)$ and $N_j(\gamma)$ 
as function of $L_j(\pi,\gamma)^{-1}=(\eta_j-1)/\eta_j$,
the curves shown in Fig. 10.

The parameters $r_3(\gamma),r_1(\gamma)$ and $N_3(\gamma),N_1(\gamma)$ 
determined in this way  converge to each other. Of course, we should have in 
mind, that our finite size analysis (\ref{sjppign}) takes into account only 
the first two terms of an infinite expansion. Approaching the isotropic limit, 
it might happen that subleading terms compete more and more with the leading 
ones. Such an effect could spoil our determination of 
$r_1(0),r_3(0),N_1(0),N_3(0)$.

In Fig. 9 the small deviations from linearity for $\gamma = 0.1\pi$ might
originate from subleading terms. The summation of these  terms might 
also change the exponent of $\ln N$ in the isotropic structure factor 
(\ref{sjppig0n}). This exponent is predicted to be 3/2 instead of 1 by 
renormalization group equations.\cite{fisher,affleck} We think, however,
that it is impossible to extract this exponent in a reliable way from the 
small systems we are investigating here. 
\section{Finite size scaling variables for all momentum values.}
The finite size effects (\ref{djpgn}) for $p<\pi$ and (\ref{sjppign})  for 
$p=\pi$ can be properly  described by the scaling ansatz
\begin{equation}\label{sasjpgn}
	S_j(p,\gamma,N)=S_j\biglb(\gamma,L_j(p,\gamma,N)\bigrb),\quad j=1,3,
\end{equation}
with a size dependent 
modification of the scaling variables (\ref{l3pg}) and (\ref{l1pg}):
\begin{eqnarray}\label{ljpgn}
	L_j(p,\gamma,N)=\frac{\eta_j}{2(\eta_j-1)}
	&&\left[	 \left(1+i\frac{a_j}{N}\right)^{\eta_j-1}
		+\left(1-i\frac{a_j}{N}\right)^{\eta_j-1}\right. \\ \nonumber
 	    &&-\left.\left(1-\frac{p}{\pi}+i\frac{a_j}{N}\right)^{\eta_j-1}-
   	\left(1-\frac{p}{\pi}-i\frac{a_j}{N}\right)^{\eta_j-1}\right].
\end{eqnarray}
 The $L_j(p,\gamma,N)$ have
the following properties:
\begin{itemize}
\item[(1)] There is no finite size effect for:
	\begin{equation}
		L_3(p,\gamma=\pi/2,N)=\frac{2}{\pi}p.
	\end{equation}
\item[(2)] For $p\neq \pi, \gamma \neq 0$ the leading finite size effects
	are of the form:
	\begin{equation}
		L_j(p,\gamma,N)-L_j(p,\gamma)=\frac{\eta_j(2-\eta_j)}{2}
		\left[1-\left(1-\frac{p}{\pi}\right)^{\eta_j-3}\right]\cdot
		\left(\frac{a_j}{N}\right)^2.
	\end{equation}
\item[(3)] For $p=\pi$ and $\gamma \neq 0$ the leading finite size behavior
	changes to:
	\begin{equation}
		L_j(p=\pi,\gamma,N)=\frac{\eta_j}{\eta_j-1}
			\left[1-\sin \left(\frac{\pi}{2}\eta_j\right)
			\left(\frac{a_j}{N}\right)^{\eta_j-1}\right].
\end{equation}
\item[(4)] For the isotropic case $\gamma=0$ (\ref{ljpgn}) reduces to:
	\begin{equation}
		L_j(p,\gamma=0,N)=\frac{1}{2} 
		\ln \frac{1+\left(a_j/N\right)^2}
		{\left(1-p/\pi\right)^2 + 
			\left(a_j/N\right)^2}.
	\end{equation}
\end{itemize} 
The quantities $a_j$ in (\ref{ljpgn}) can still depend on $p$ and $\gamma$. 
For $p<\pi, \gamma >0 $, they are related 
to the coefficients $c_j(p,\gamma)$ describing the
finite size behavior (\ref{djpgn}):
\begin{equation}
	c_j(p,\gamma)=\frac{\partial S_j}{\partial L_j} \cdot
		\frac{\eta_j(2-\eta_j)}{2} \cdot
		\left[1-\left(1-\frac{p}{\pi}\right)^{\eta_j-3}\right]
		a_j(p,\gamma)^2.
\end{equation}
The derivative of the scaling functions $S_j(\gamma,L_j)$ with
respect to the scaling variable $L_j$ is close to one up to the small
deviations coming from the difference $\Delta_j(p,\gamma)$ depicted in
Figs. 7(a)-(f). Note that the size dependent scaling variable 
(\ref{ljpgn}) describes correctly the singularity structure (\ref{varfig6}) of 
$c_j(p,\gamma)$ for $p\rightarrow \pi, \gamma >0 $. Therefore the quantities 
$a_j(p,\gamma)$ are expected to be free of singularities for $p=\pi$.\\
Under the premise that the scaling ansatz (\ref{sasjpgn}) is correct and the
finite size behavior at $p=\pi$ is given by (\ref{sjppign}), the scaling
function in (\ref{sasjpgn}) has to increase linearly with the scaling variable
(\ref{ljpgn}) for large values of $L_j$. In particular, we find for the
diverging transverse structure factor at $p=\pi$ and
$N\rightarrow \infty$:
\begin{equation}
	\left.S_1\biglb(\gamma,L_1(p,\gamma,N)\bigrb)\right|_{p=\pi}
		\stackrel{N\to\infty}{\longrightarrow}
		 r_1(\gamma)\cdot \frac{\eta_1}{\eta_1 -1}
		\left[1 - \sin\left(\frac{\pi}{2}\eta_1\right) 
		\left(\frac{a_1}{N}\right)^{\eta_1-1}\right].
\end{equation}
Comparing with (\ref{sjppign}) we see that the scaling ansatz (\ref{sasjpgn}) 
with the 
scaling variable (\ref{ljpgn}) provides us as well with the singular behavior
of the coefficient on the right hand side of (\ref{sjppign}) in the isotropic 
limit $\gamma \rightarrow 0 $. Moreover there is a relation between 
$N_1(\gamma)$ and $a_1(p=\pi,\gamma)$:
\begin{equation}\label{n1g}
	N_1(\gamma)=a_1(p=\pi,\gamma) \cdot
	\left(\sin\frac{\pi}{2}\eta_1\right)^{-\pi/\gamma}.
\end{equation}
In contrast to the transverse structure factor, the longitudinal one
stays finite for $p=\pi,N\rightarrow \infty,\gamma>0$. The comparison
of (\ref{sjppign}) with the scaling ansatz (\ref{sasjpgn}) and (\ref{ljpgn}) 
yields here:
\begin{equation}
	\left.S_3\biglb(\gamma,L_3(p,\gamma)\bigrb)\right|_{p=\pi}
		=r_3(\gamma)\frac{\eta_3}{\eta_3-1},
\end{equation}
\begin{equation}\label{n3g}
	N_3(\gamma)=a_3(p=\pi,\gamma)\cdot
	\left[\frac{\partial \ln S_3}{\partial L_3}
	\cdot\frac{\eta_3}{\eta_3-1}
	\sin \left(\frac{\pi}{2}\eta_3\right)\right]^{\pi/\gamma-1}.
\end{equation}   
Again we find the pole in the coefficient on the right hand side of
(\ref{sjppign}), provided that the scaling function $S_3(\gamma,L_3)$ increases
linearly with $L_3(p=\pi,\gamma)$ [cf.(\ref{d3pgn})]. The linear behavior also 
guarantees that $N_1(\gamma)$ [cf.(\ref{n1g})] and
$N_3(\gamma)$ [cf.(\ref{n3g})] converge to the same value if:
\begin{equation}
	a_1(p=\pi,\gamma=0)=a_3(p=\pi,\gamma=0).
\end{equation}
Summarizing we can say:
The scaling ansatz (\ref{sasjpgn}) with the scaling variable (\ref{ljpgn}) 
reproduces correctly:
\begin{itemize}
\item[(1)] The finite size behavior (\ref{djpgn}) at $p<\pi$ and 
	(\ref{sjppign}) at $p=\pi$.
\item[(2)] The singular behavior (\ref{varfig6}) of the coefficients 
	$c_j(p,\gamma)$ for $p\rightarrow \pi$.
\item[(3)] The singular behavior of the coefficients on the right hand side
	of (\ref{sjppign}) for $\gamma \rightarrow 0$.
\end{itemize}
On the other hand, the quantities $a_j(p,\gamma)$, which enter in the
definition of the scaling variable (\ref{ljpgn}) appear to be free of 
singularities at $p=\pi$ and $\gamma=0$.\\
Finally, we would like to mention that the term $a_jN^{-1}$ in (\ref{ljpgn})
can be interpreted as an inverse correlation length $\xi_j(\gamma,N)^{-1}$.
The Fourier transform of (\ref{ljpgn}) which brings us back to the 
configuration space (with coordinate $x$) has a large distance behavior:
\begin{equation}
	\frac{(-1)^x}{x^{\eta_j}}\exp\left(-\frac{x}{\xi_j(\gamma,N)}\right).
\end{equation}
The effective correlation length $\xi_j(\gamma,N)=N/a_j$ takes into account
the finiteness of the system. In the critical regime $0\leq\gamma\leq\pi /2$
---we are considering here--- the effective correlation length is assumed to
increase with the system size $N$.
\section{Conclusion}
In this paper we have analysed the longitudinal and transverse structure 
factors $S_j(p,\gamma,N)$ of the XXZ model on finite rings $N \leq 28$. Based
on this analysis we got a reliable estimate of the thermodynamical limits 
$N\to\infty$ for momenta $p \leq 13\pi/14$ and for 
$ \gamma/\pi = 0,0.1,0.2,0.3,0.4,0.5$. We observe two kinds of structures:
\begin{itemize}
\item[(1)] A ``gross''-structure, which is given by the universal 
		scaling variables
  	(\ref{l3pg}) and (\ref{l1pg}).
\item[(2)] A ``fine''- structure, which contributes only a few percents and 
	which depends explicitly on $\gamma$. In particular we found a 
	minimum in the longitudinal structure factors at momentum $p=p(\gamma)$
	listed in TABLE I. A comparison of our finite size analysis with the 
	exact result \cite{tonegawa} on the structure factors in the XX model 
	($\gamma=\pi/2$) yields excellent agreement.

	A comparison with the conjecture of M\"uller\cite{mueller} {\it et al.}
	reveals systematic deviations for the transverse structure factors 
	and striking discrepancies for the longitudinal structure factors.
\end{itemize}
Exact diagonalizations of the Hamiltonian -- as they were performed in this
paper for the XXZ-model -- are restricted to rather small systems 
($N \lesssim 36$) in contrast to quantum Monte Carlo simulations which are
feasible on much larger systems ($N\lesssim 1000$). Monte Carlo simulations,
however, suffer under statistical errors, which simply prevent the resolution 
of the ``fine''-structures mentioned above.

On the other hand these structures are clearly visible in exact calculations on
small systems and we are sure that they survive in the thermodynamical limit
for the following reason. The finite size effects of the structure factors
$S_j(p,\gamma,N),j=1,3$ are well under control, which we demonstrated by
comparison with the exact result for the XX case ($\gamma=\pi/2$). It should be
possible to extend this type of finite size analysis to the dynamical structure
factors. Work along this line is in progress.
\begin{appendix}
\section{}
We are going to prove that the longitudinal structure factors for the
groundstate on a ring with $N$ sites is given by:
\begin{equation}\label{s34kn}
	S_3\left(p=2\pi k/N,\gamma=\pi/2,N\right) = \frac{4k}{N}, 
	\qquad k=0,1,...,\frac{N}{2},
\end{equation}
which means for spin-spin correlators in the groundstate $|0\rangle$:
\begin{eqnarray}
	4\langle 0 |S_3(0)S_3(x) |0 \rangle(\gamma=\pi/2,N) = 
		\left\{ \begin{array}{cl} 1 & \text{ for } x = 0 \\
				 0 & \text{ for }  x =2l \\
	          -\frac{4}{N^2\sin^2\left( \pi x/N \right)} & 
				\text{ for }  x = 2l+1.
		\end{array}\right.
\end{eqnarray}
Following Lieb, Schultz and Mattis\cite{lsm} the spin-spin correlators
 can be represented as:
\begin{eqnarray}\label{lsms0sl}
	4\langle 0 |S_3(0)S_3(l) |0 \rangle (\gamma=\pi/2,N) = 
	(G_0^\sigma)^2 - (G_l^\sigma)^2,
\end{eqnarray}
where
\begin{eqnarray}\label{gls}
	G_l^\sigma = -\sum_{k=0}^{N-1} \Phi_{k0}^\sigma \Phi_{kl}^\sigma 
		{\rm sgn}  \left(\Lambda_{k}^\sigma\right).
\end{eqnarray}
The $\Phi_{kl}^\sigma$ are solutions of the equations:

\begin{mathletters}\label{eigenvalue}
\begin{eqnarray}
	 \Phi_{k2}^\sigma + \sigma \Phi_{kN}^\sigma 
		&=& \Lambda_k^\sigma \Phi_{k1}^\sigma, \\
         \Phi_{k,l-1}^\sigma+\Phi_{k,l+1}^\sigma 
		&=& \Lambda_k^\sigma\Phi_{kl}^\sigma,\\
        \sigma\Phi_{k1}^\sigma + \Phi_{k,N-1}^\sigma  
		&=& \Lambda_k^\sigma \Phi_{kN}^\sigma. 
\end{eqnarray}
\end{mathletters}
These equations are obtained by the diagonalization of the XX Hamiltonian by
means of a Jordan-Wigner-Transformation. In the derivation of these equations
one has to distinguish between the chains $N=4,8,10,...;(\sigma=-1)$ and 
$N=6,10,14,...;(\sigma=+1)$. Indeed equations (\ref{eigenvalue}) were derived 
by Lieb, Schultz and Mattis \cite{lsm} for $\sigma=1$. The derivation for the 
second case is straight forward. It is easy to verify that the equations
(\ref{eigenvalue}) are solved by:
\begin{mathletters}
\begin{eqnarray}
	\Phi_{kl}^+ &=& \sqrt{\frac{2}{N}}\left\{     \begin{array}{ll}
	1/\sqrt{2} & \text{ for } k = 0 \\
        \cos\left(2kl\pi/N\right)
		& \text{ for } k = -1,-2,...,-N/2+1 \\
	\sin\, \left(2kl\pi/N\right)
		&  \text{ for } k = +1,2,...,N/2-1 \\
	(-1)^l/\sqrt{2} & \text{ for } k = N/2, 
		                     \end{array}\right. \\
	\Phi_{kl}^- &=& \sqrt{\frac{2}{N}}\left\{     \begin{array}{ll}
        \cos [ \left(2k+1\right)l\pi/N]
		& \text{ for } k = 0,-1,...,-N/2+1 \\
	\sin\, [\left(2k+1\right)l\pi/N]
		&  \text{ for } k = +1,2,...,N/2, 
		                     \end{array}\right. \\
	\Lambda_k^\sigma &=& 
	\cos\left[\frac{\pi}{N}\left(2k+\frac{1}{2}(1-\sigma)\right)\right],
	\qquad k = -\frac{N}{2}+1,-\frac{N}{2},...,\frac{N}{2}.
\end{eqnarray}
\end{mathletters}
Our proof of (\ref{s34kn}) is completed if we insert these solutions into 
(\ref{gls}) and (\ref{lsms0sl}).
\end{appendix}
{\bf FIGURE CAPTIONS}
\begin{itemize}
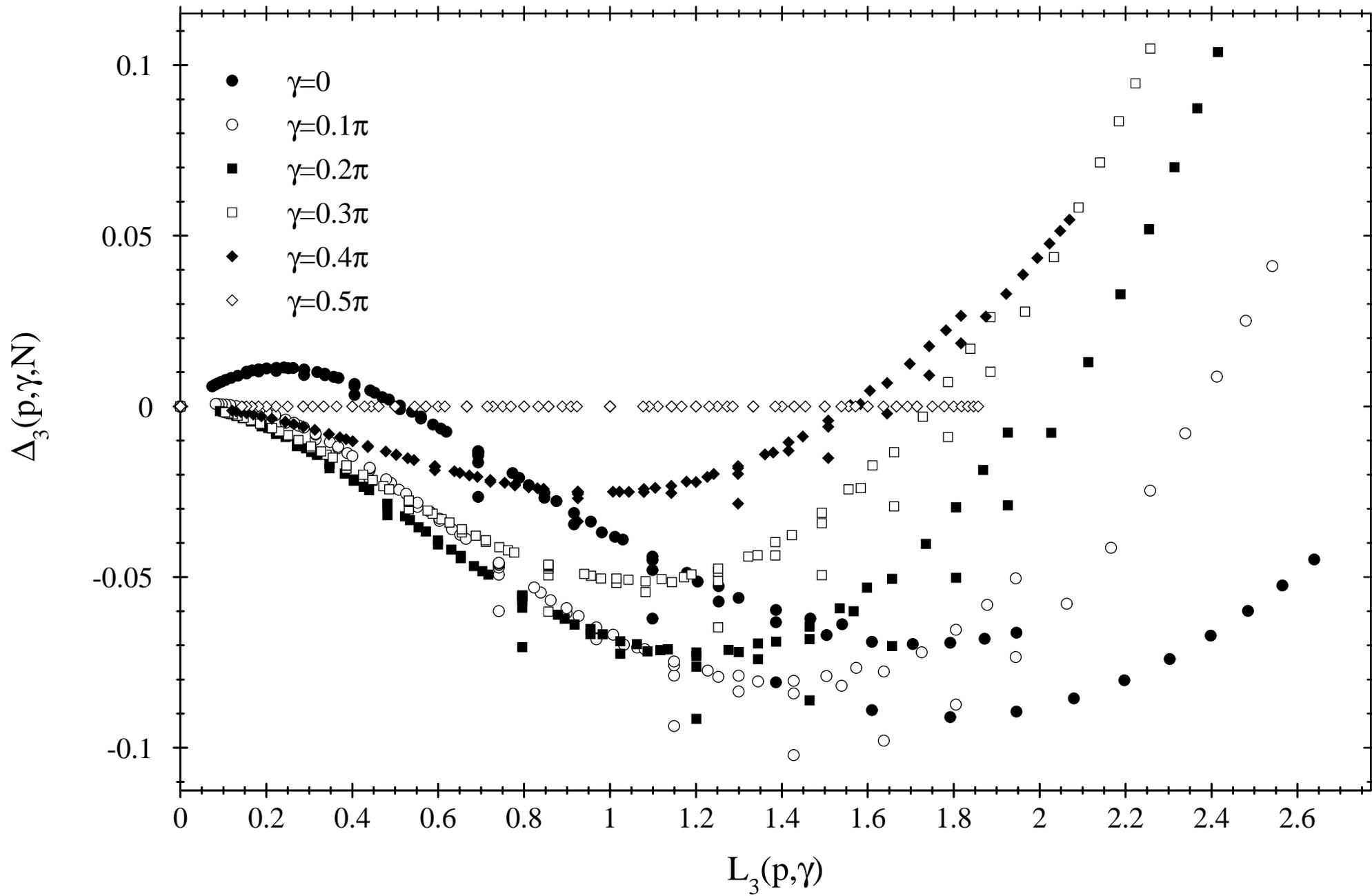
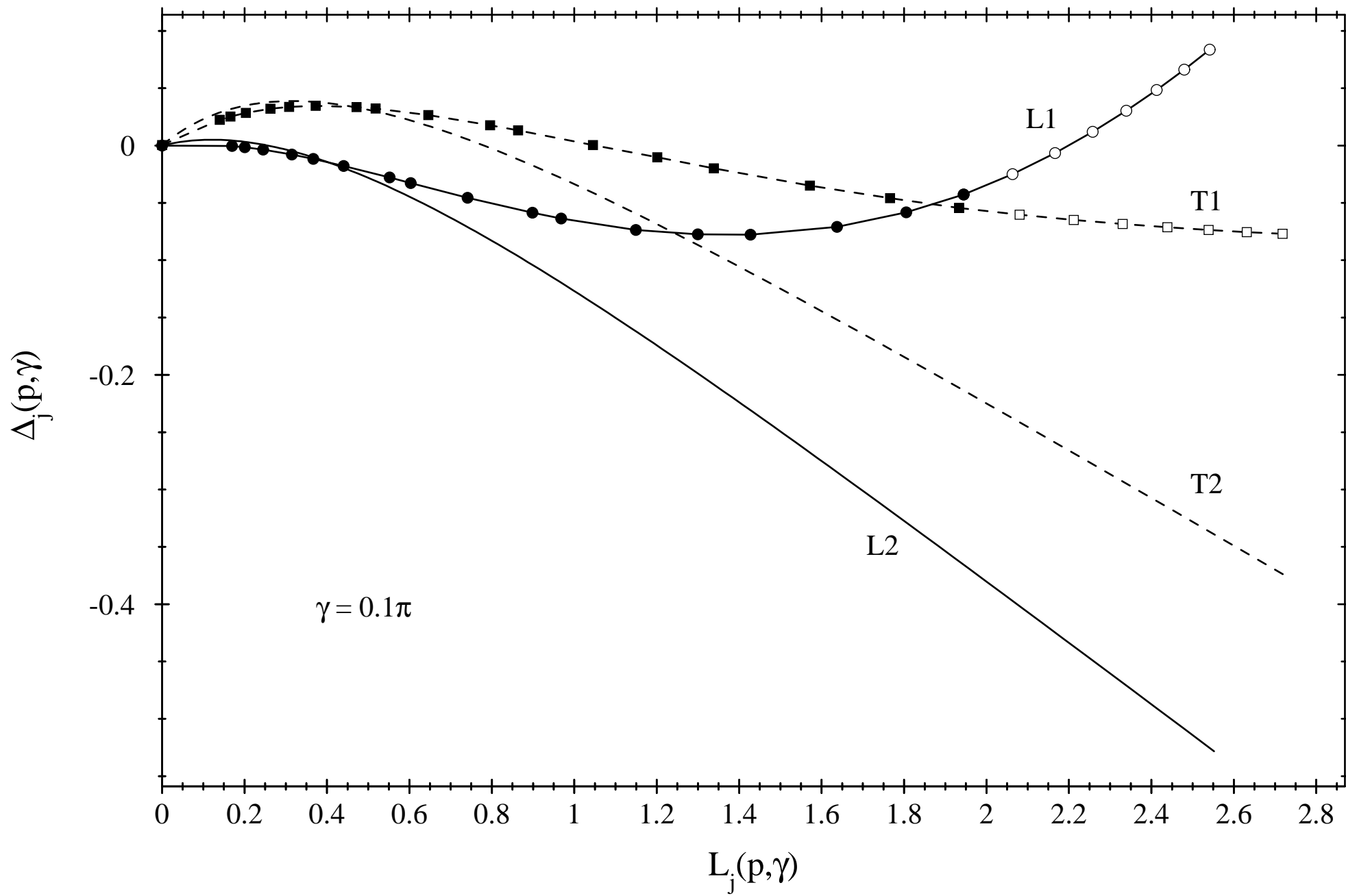
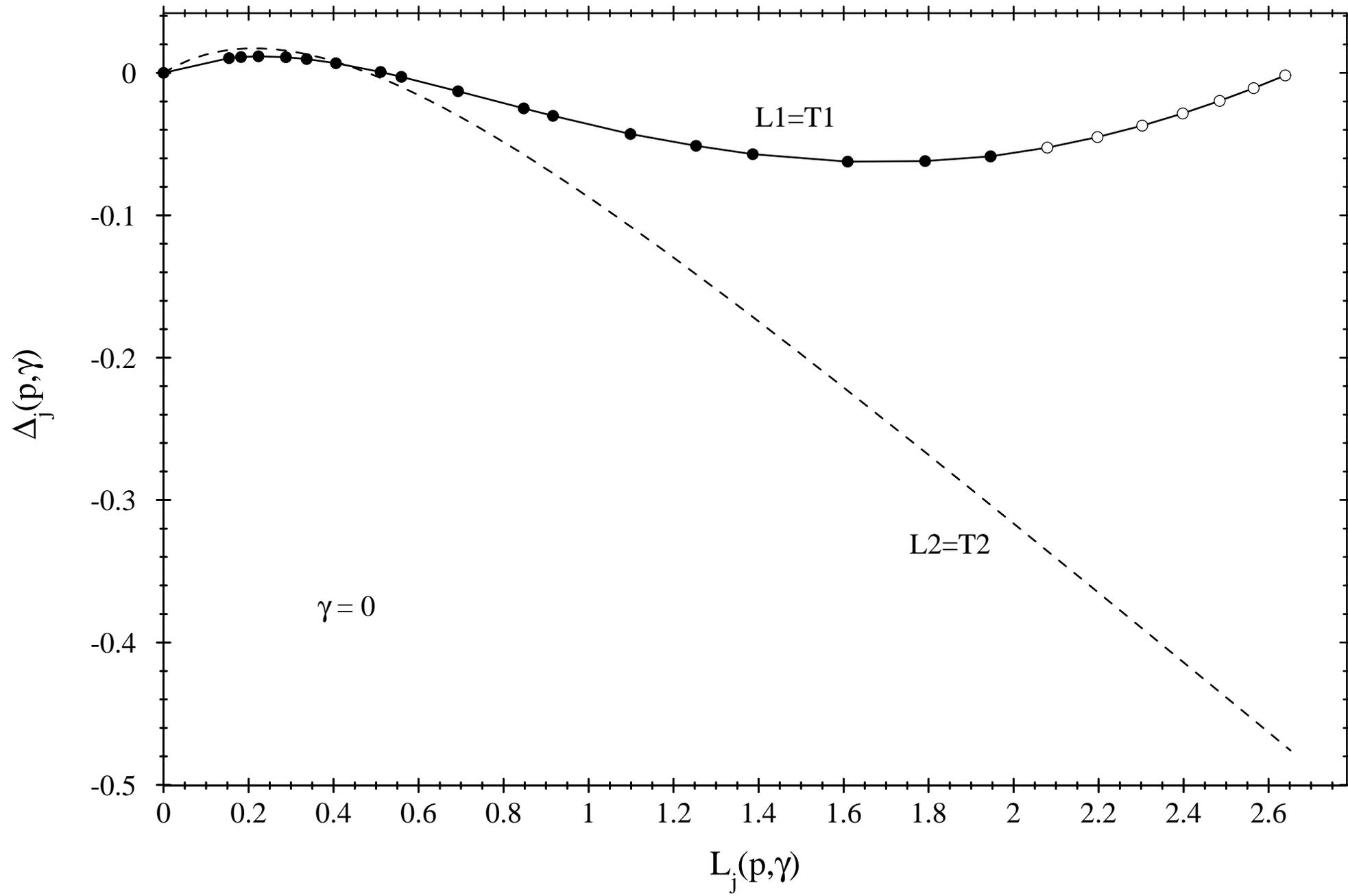
\item[FIG. 1.] The longitudinal structure factor  versus the scaling
	variable (\ref{l3pg}) for $\gamma/\pi = 0, 0.1, 0.2, 0.3, 0.4, 0.5$
	and $N = 4, 6,....28$.
\item[FIG. 2.] The difference (\ref{d3pgn})  versus the scaling
	variable (\ref{l3pg}).
\item[FIG. 3.] Same as Fig. 1 for the transverse structure factor versus the 
	scaling
	variable (\ref{l1pg}).
\item[FIG. 4.] The difference (\ref{d1pgn})  versus the scaling
	variable (\ref{l1pg}).
\item[FIG. 5.] Finite size analysis of $\Delta_j(p,\gamma,N)$ at momentum 
		$p = \pi/2$:
	\begin{itemize}
		\item[(a)] for the longitudinal structure factor,
		\item[(b)] for the transverse  structure factor.
	\end{itemize}
\item[FIG. 6.] The coefficient $c_j(p,\gamma), j=1,3$ in the finite size 
	ansatz (\ref{djpgn}) as a function of (\ref{varfig6}).
\item[FIG. 7.] The thermodynamical limit $N \to \infty$ of the transverse 
	structure
	factor:
	\begin{itemize}
		\item[T1:] our finite size analysis
		\item[T2:] conjecture Ref. \onlinecite{mueller}
		\item[T3:] exact result Ref. \onlinecite{tonegawa} at
			 $\gamma = \pi/2$
	\end{itemize}
	and of the longitudinal structure factor:
	\begin{itemize}
		\item[L1:] our finite size analysis
		\item[L2:] conjecture Ref. \onlinecite{mueller}
	\end{itemize}
	(a) $\gamma=0.5\pi$, (b) $\gamma=0.4\pi$, (c) $\gamma=0.3\pi$, 
	(d) $\gamma=0.2\pi$, (e) $\gamma=0.1\pi$, (f) $\gamma=0$.
\item[FIG. 8.] The scaling function $g_j(z=2,\gamma), j=1,3$ in the finite 
	size scaling ansatz (\ref{sa}) as a function of $\gamma$.
\item[FIG. 9.] Large $N$ behavior of the structure factors (\ref{sjppign}) at 
	the critical momentum $p =\pi$.
\item [FIG. 10.] The coefficients $r_j(\gamma), N_j(\gamma)\quad j=1,3$ in the 
	finite size ansatz (\ref{sjppign}) as a function of 
	$L_j(p,\gamma)^{-1}$,\\
		(a) $r_j(\gamma)$, (b) $N_j(\gamma)$.

\end{itemize}

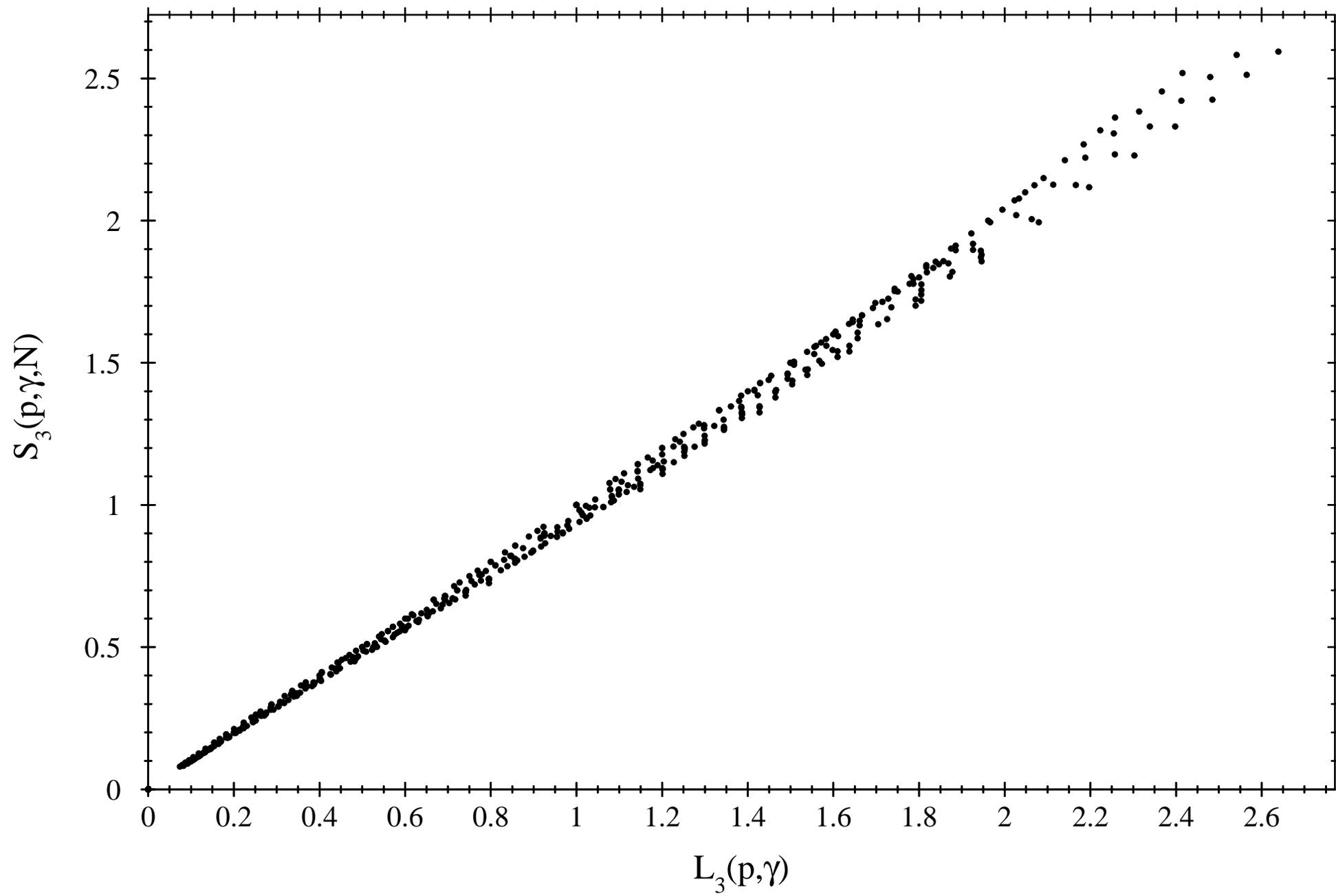
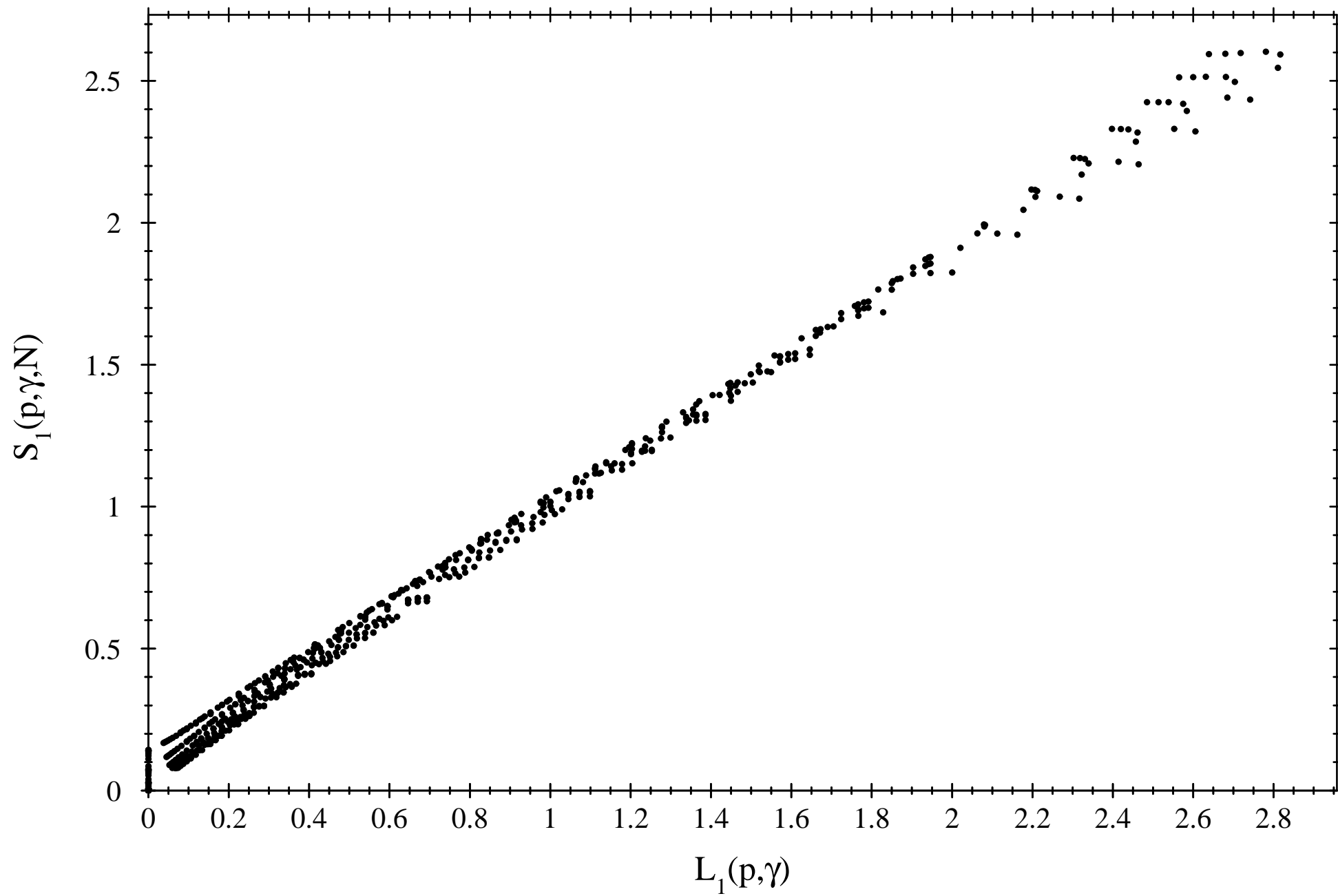
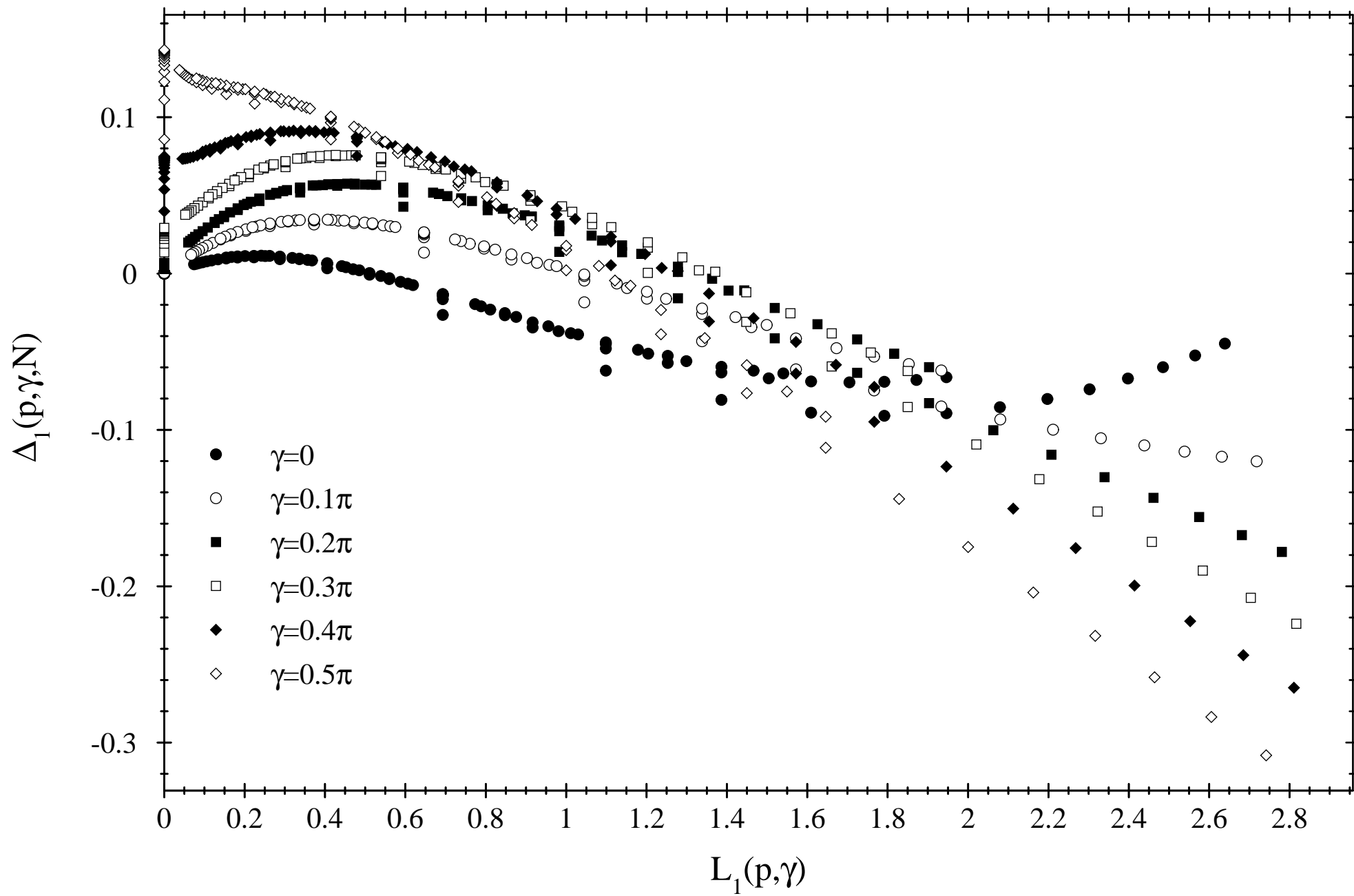
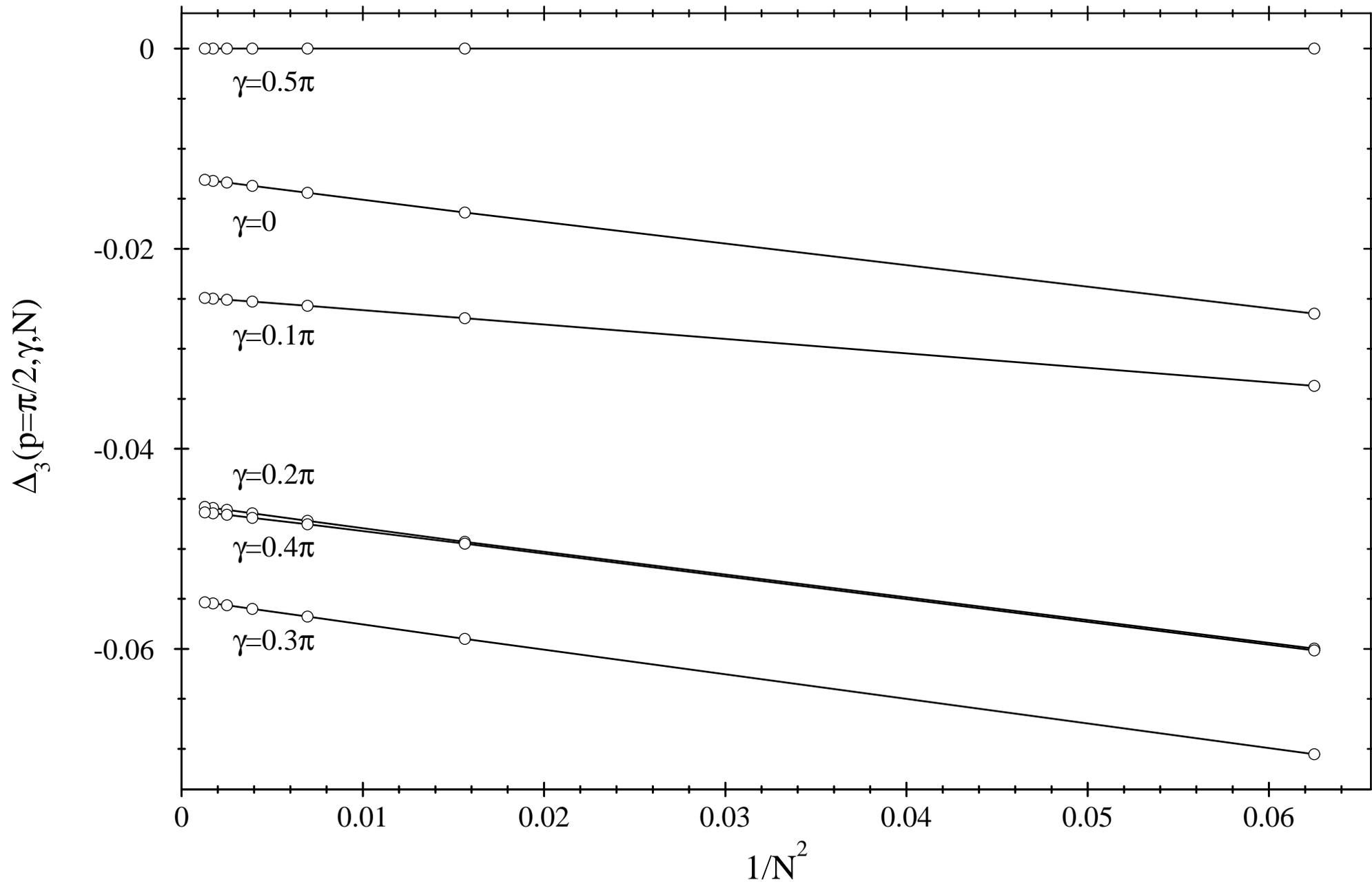
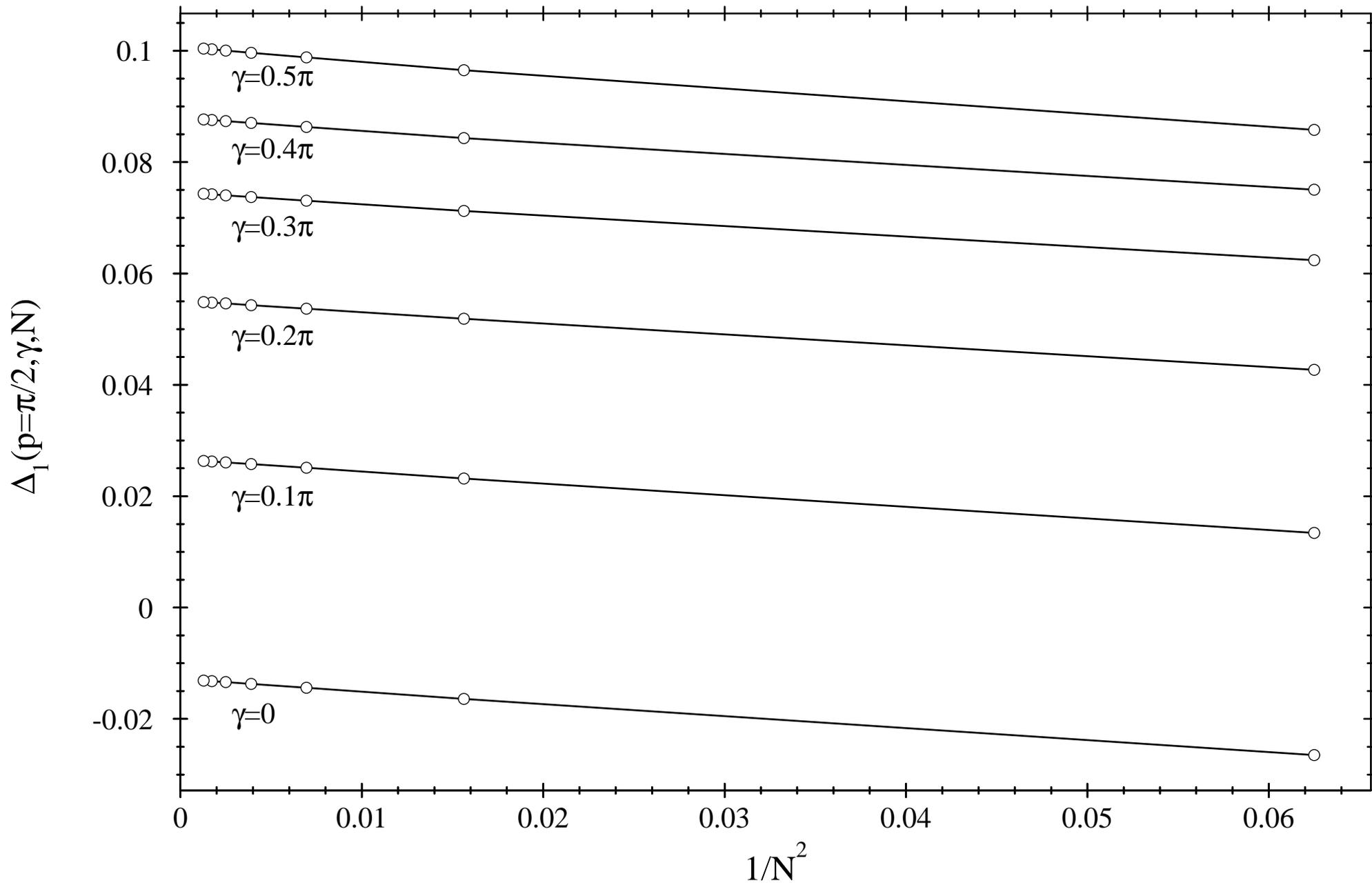
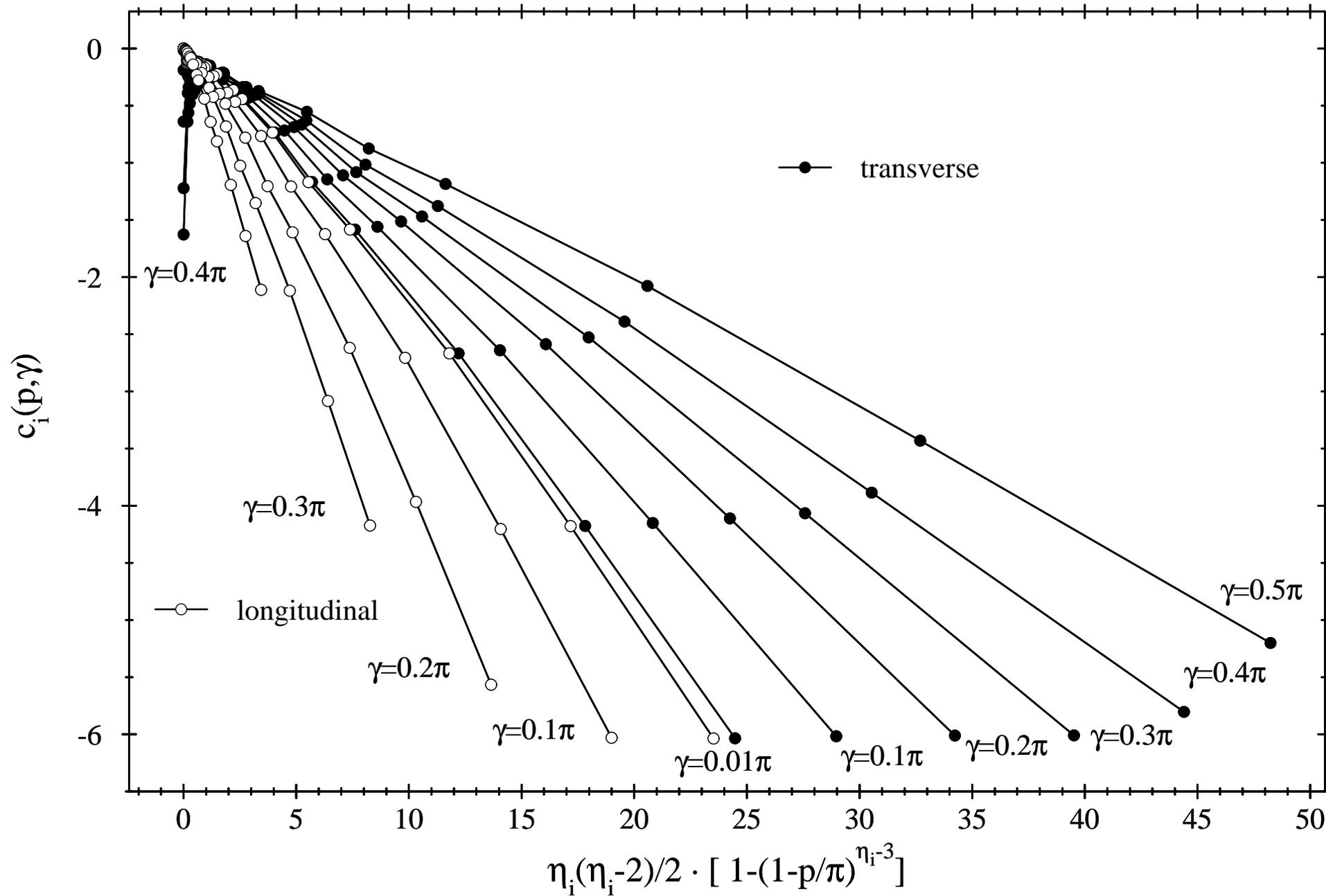
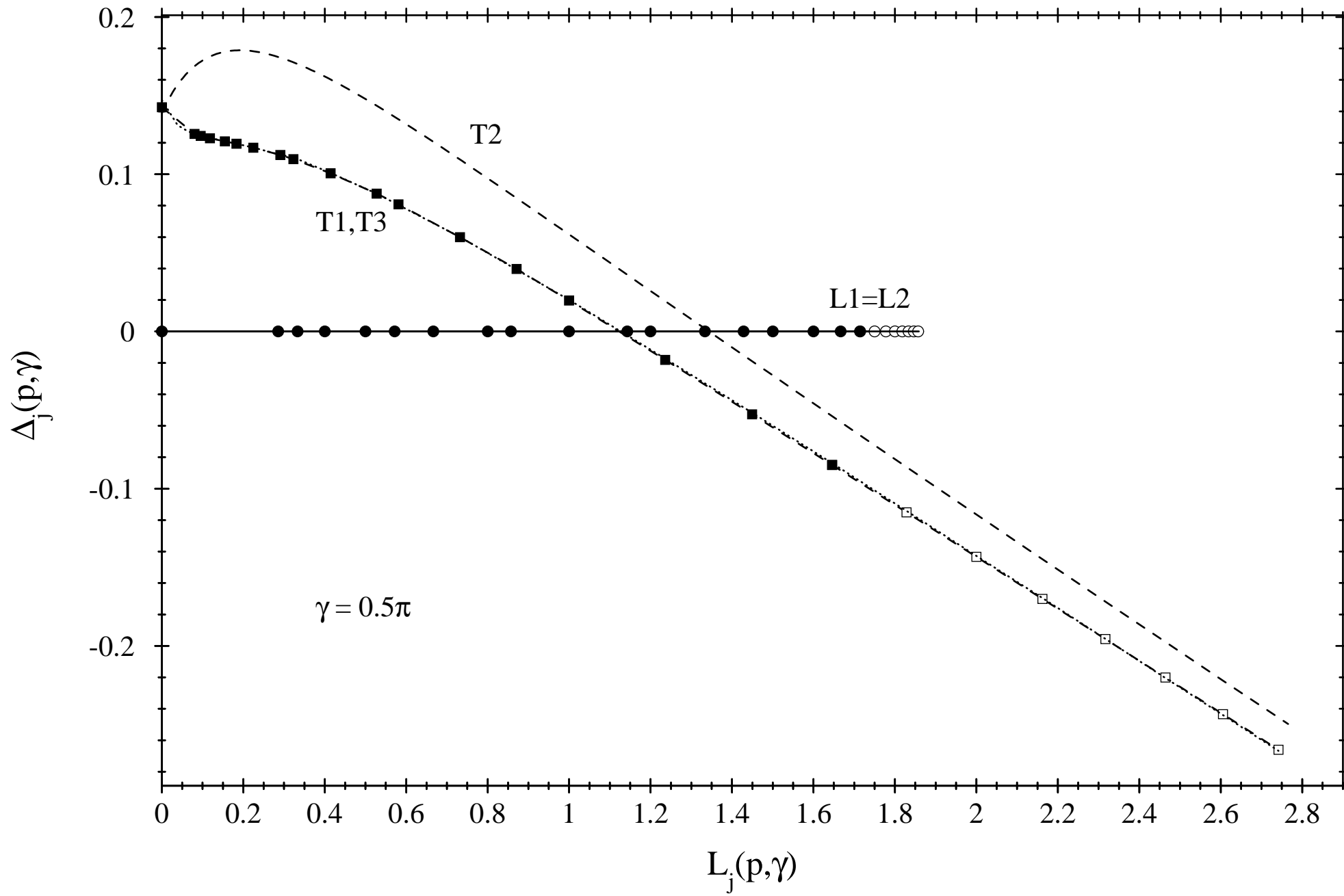
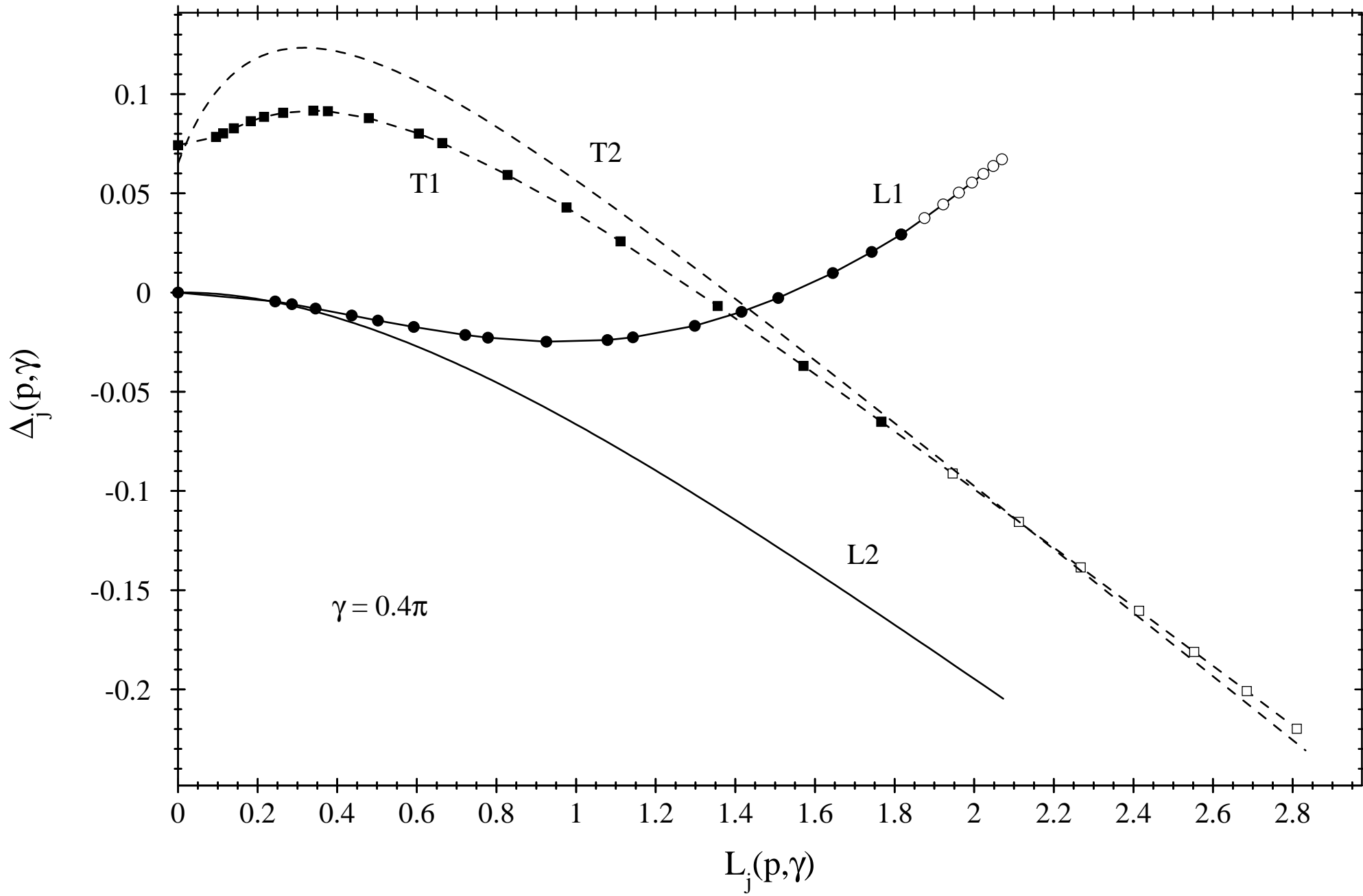
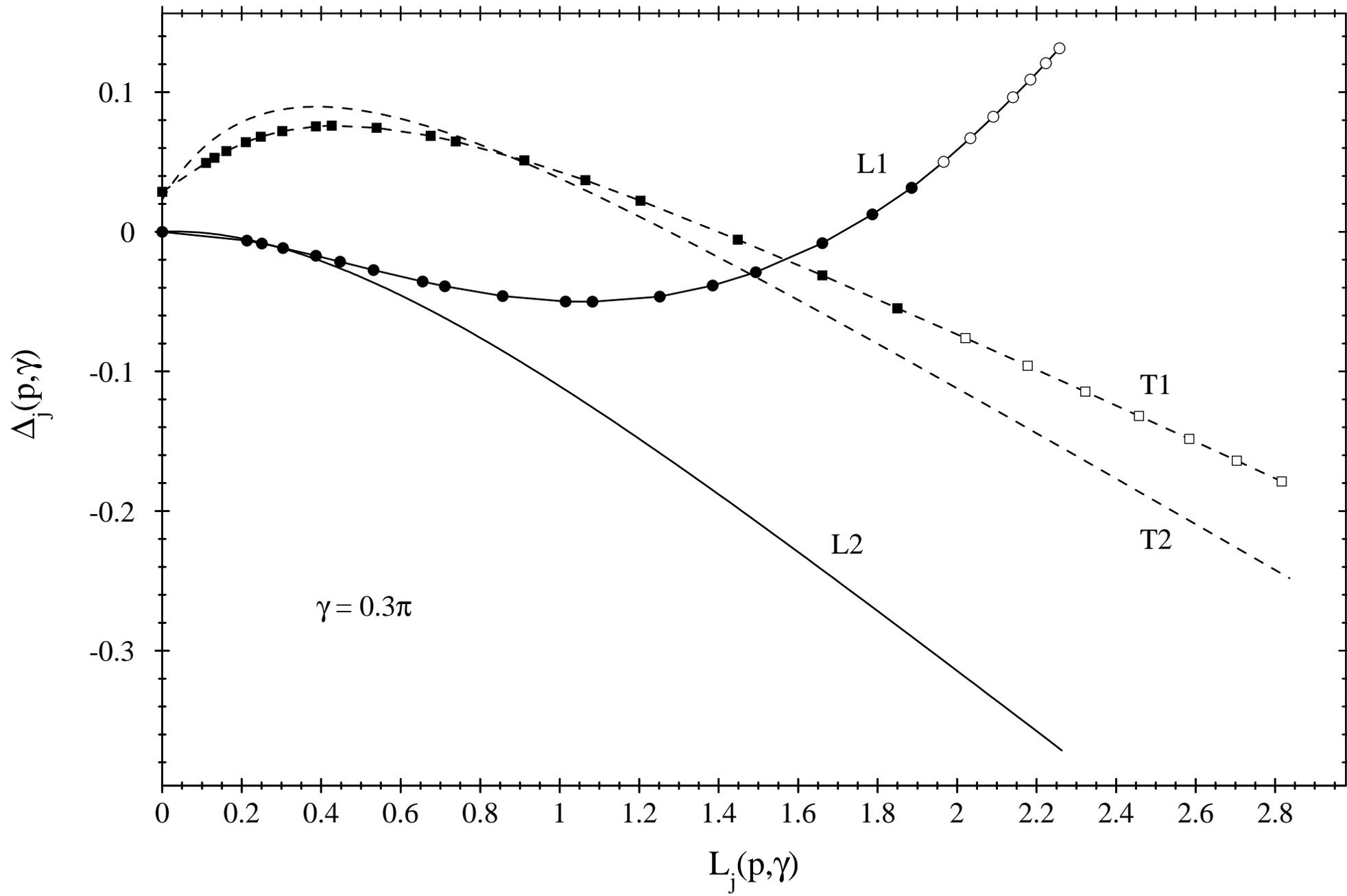
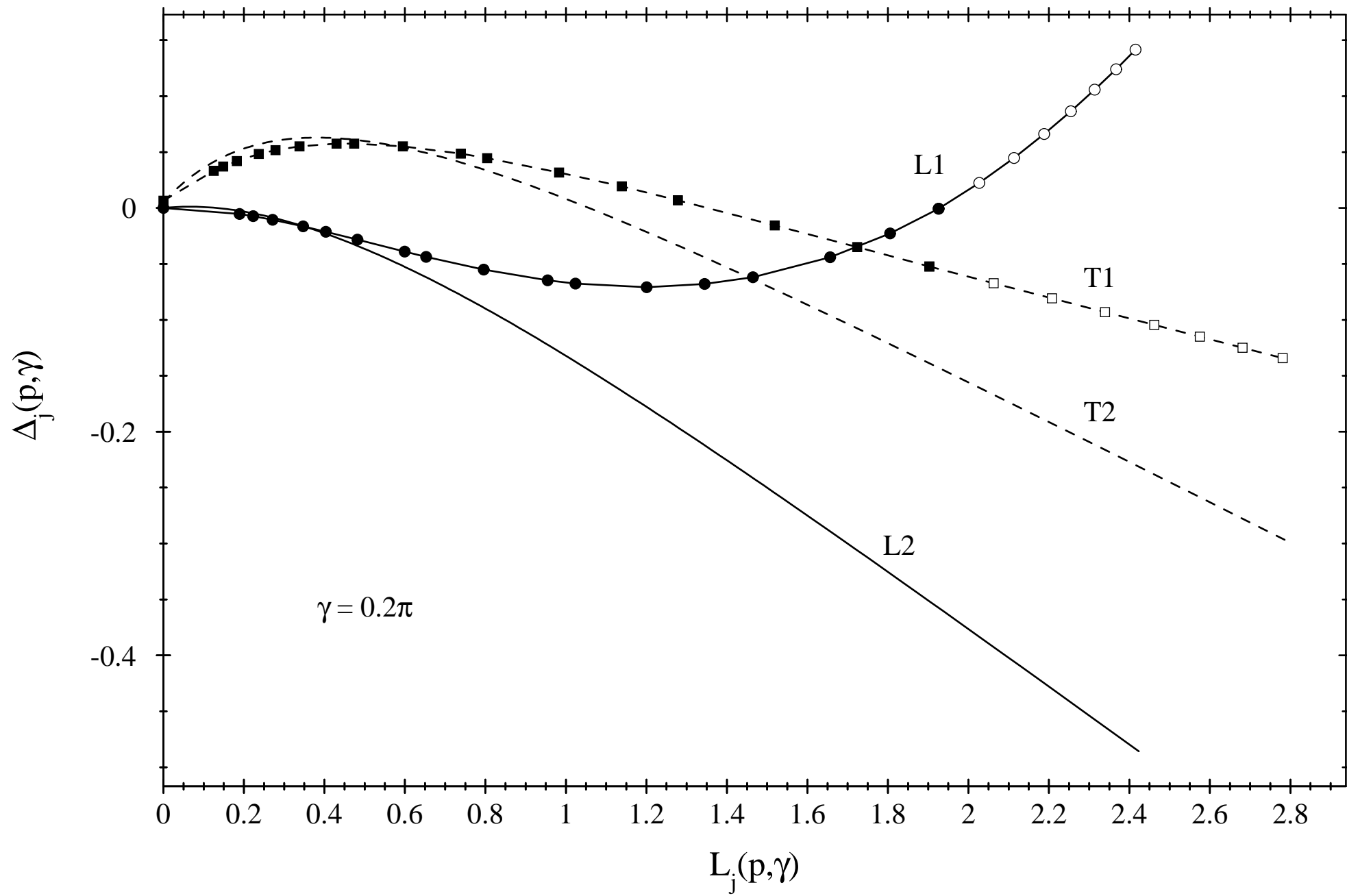
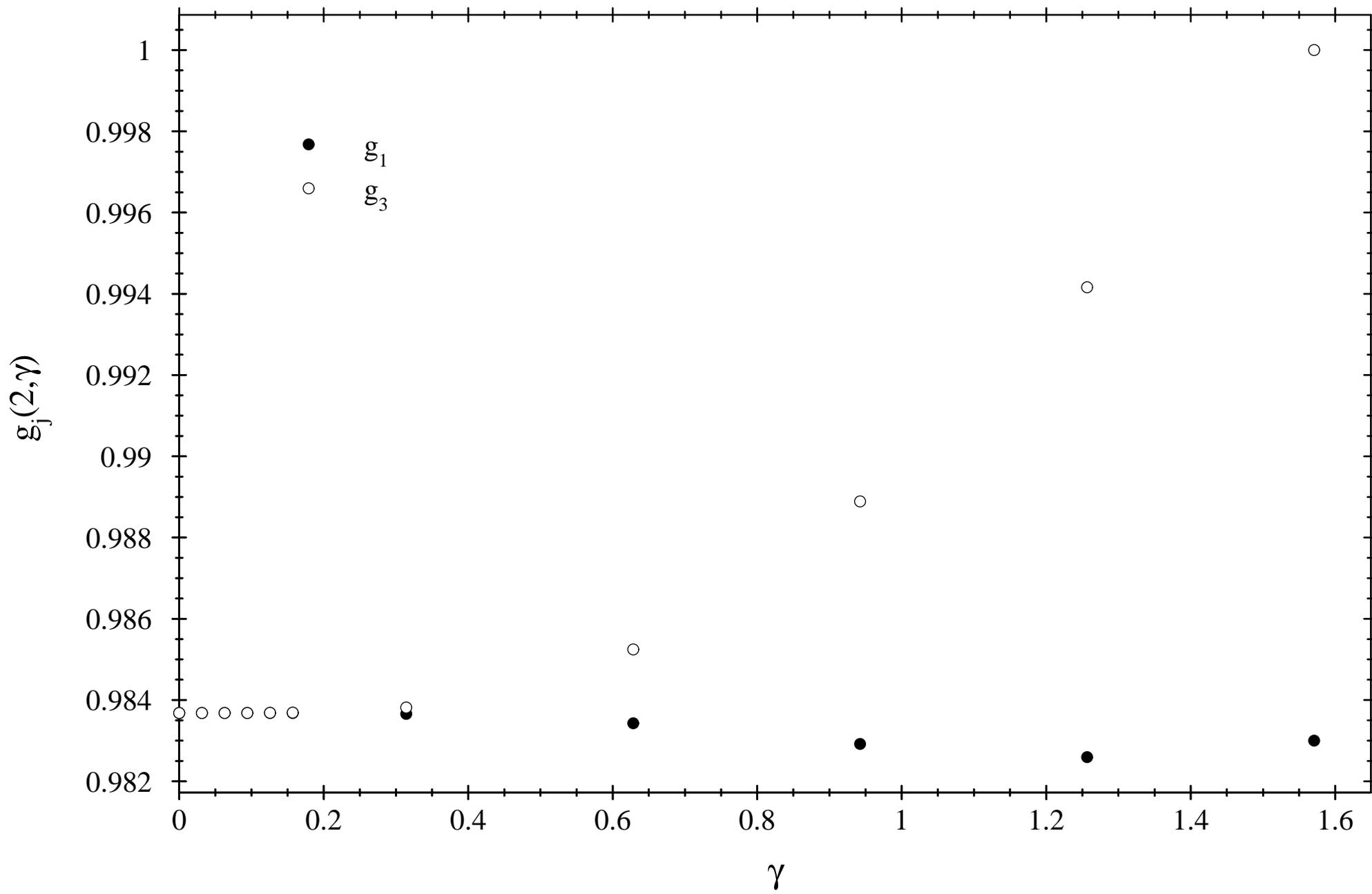
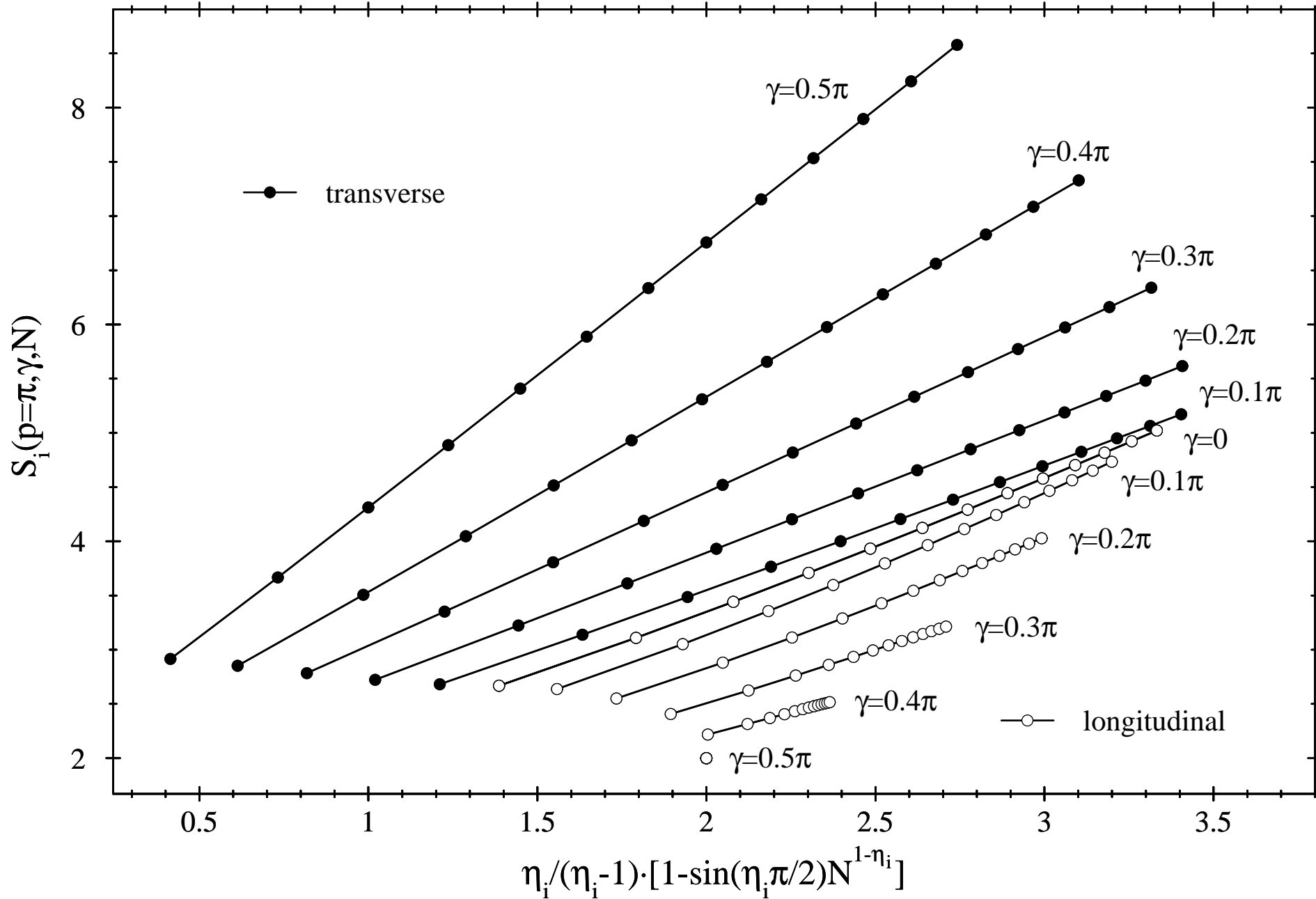
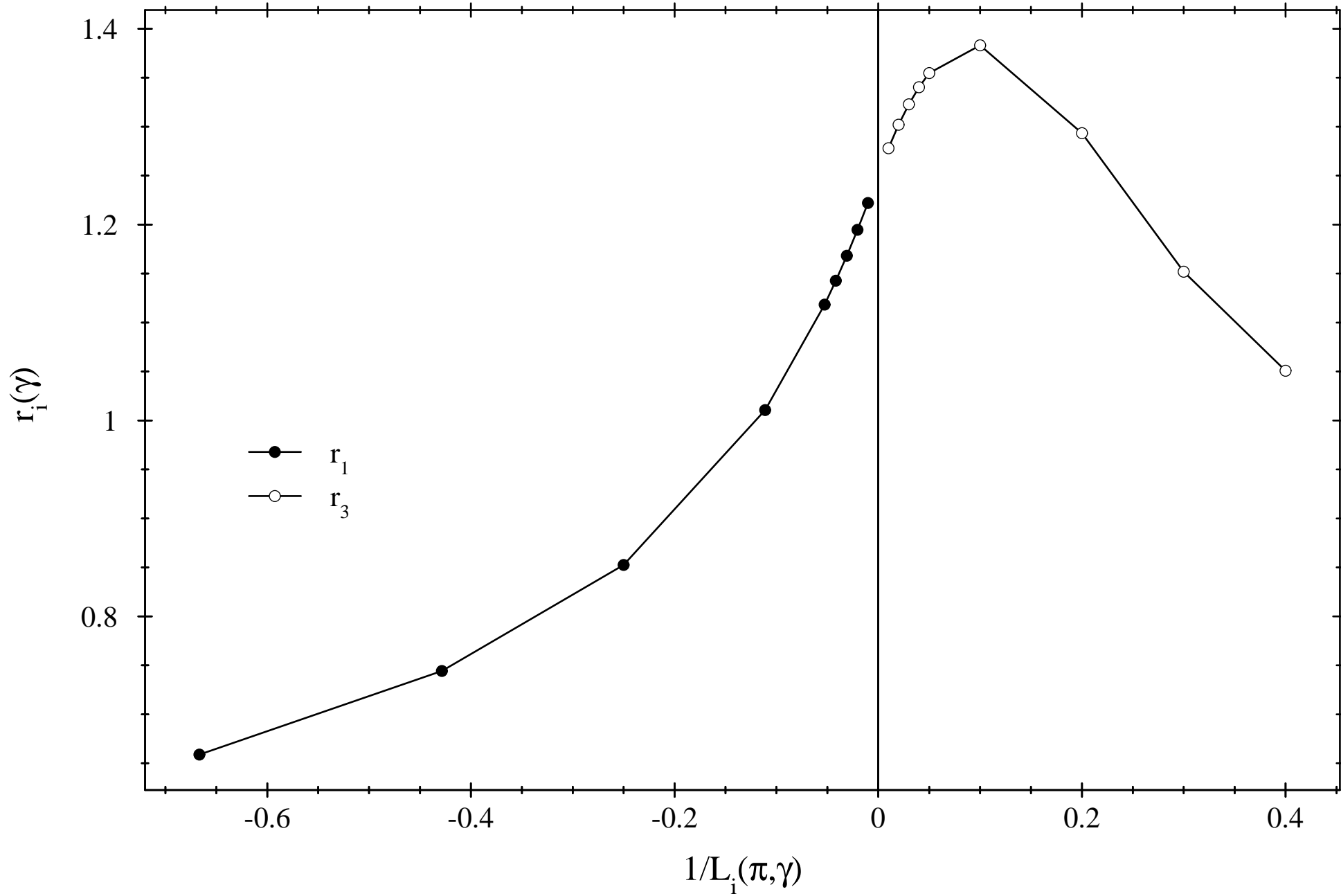
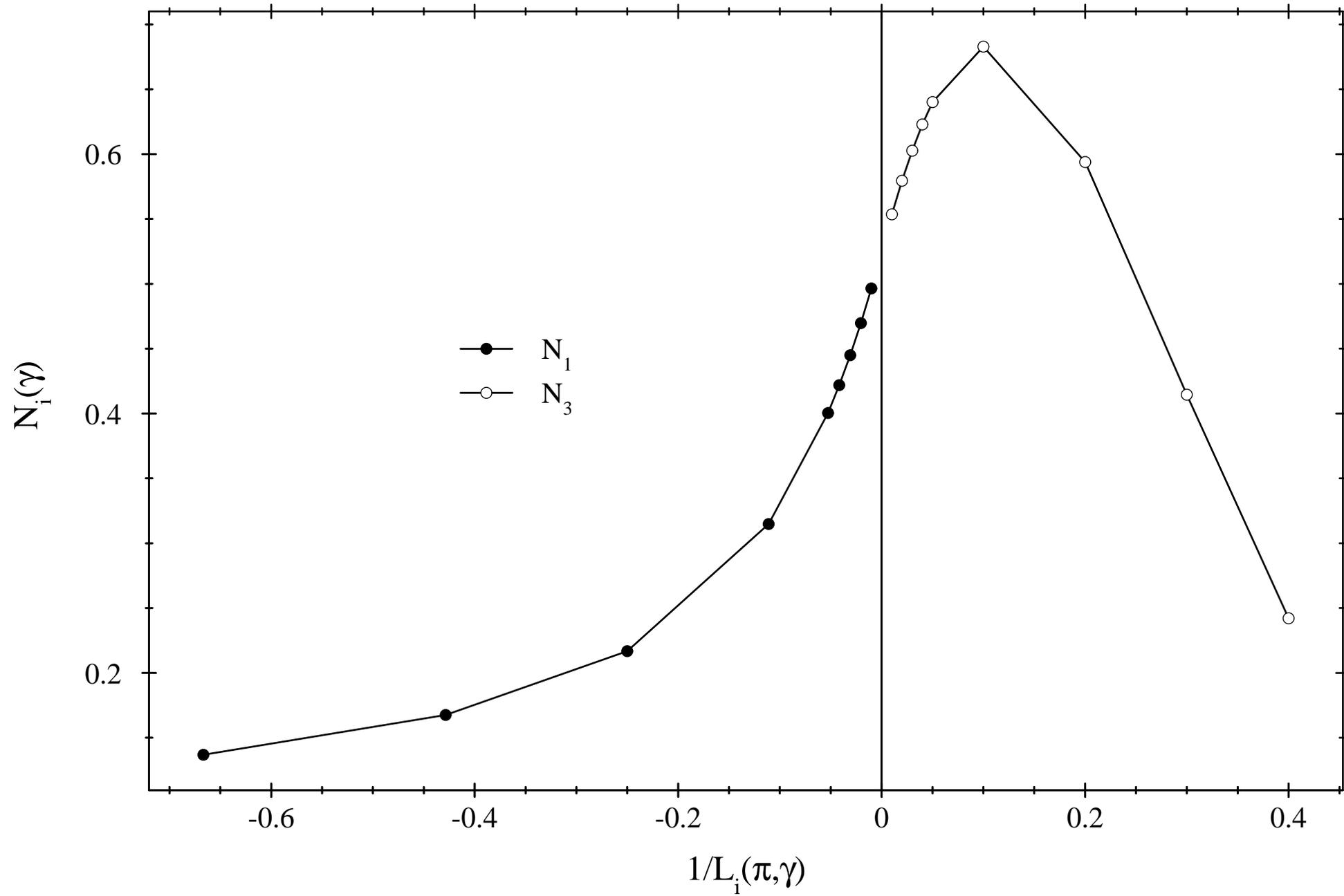
\end{document}